\documentclass[twocolumn,aps,amssymb,,superscriptaddress]{revtex4}
\usepackage{graphicx}

\begin{document}
\title{Terahertz spectroscopy of electromagnons in Eu$_{1-x}$Y$_x$MnO$_3$
}
\author{A. Pimenov}
\affiliation{Experimentelle Physik IV, Universit\"{a}t W\"{u}rzburg,
97074 W\"{u}rzburg, Germany} %
\author{ A. Loidl}
\affiliation{Experimentalphysik V, EKM, University of Augsburg,
86135 Augsburg, Germany} %
\author{A.$\>$A. Mukhin}
\author{V.D. Travkin}
\author{V.Yu. Ivanov}
\affiliation{General Physics Institute of the Russian Acad. of
Sciences, 119991 Moscow, Russia}%
\author{A. M. Balbashov}
\affiliation{Moscow Power Engineering Institute, 105835 Moscow,
Russia}
\date{\today}

\begin{abstract}
Dielectric permittivity spectra of yttrium-doped EuMnO$_3$ in the
composition range $0 \le x \le 0.5$ have been investigated in the
terahertz frequency range. Magnetoelectric contributions to the
permittivity were observed in all compositions for ac electric
fields parallel to the crystallographic a-axis. Well defined
electromagnons exist for $x \geq 0.2$ close to $\nu \sim 20$
cm$^{-1}$ and with dielectric strength strongly increasing on
doping. In addition to electromagnons, a broad contribution of
magnetoelectric origin is observed for all compositions. For
Eu$_{0.8}$Y$_{0.2}$MnO$_3$ the electromagnons can be suppressed by
external magnetic fields which induce a canted antiferromagnetic
phase. Magnetoelectric effects in the different doping regimes are
discussed in detail.

\end{abstract}

\pacs{75.80.+q,75.47.Lx,75.30.Ds}

\maketitle

\section{Introduction}

Electric and magnetic properties in most cases are well separated in
the physics of the solid state. However, as can be expected already
from the Maxwell equations, the cross-coupling effects between
electricity and magnetism may exist under appropriate symmetry
conditions. This cross-coupling has been pointed out already by
Pierre Curie \cite{curie} and was later called magnetoelectric (ME)
effect. ME effects can be observed e.g. as changes in the electric
polarization in external magnetic fields or as electric-field
dependence of the magnetic moment. Among the systems revealing the
ME effect special interest is drawn to materials with the
simultaneous occurrence of (anti-)ferromagnetism and
ferroelectricity, which are termed multiferroics \cite{schmid}. In
addition to unusual physical properties the multiferroics are also
attractive from the point of view of possible applications
\cite{hill,fiebig,khomskii}, e.g. as novel memory elements and
optical switches. Multiferroic behavior occurs in a variety of
systems originating from very different physical mechanisms,
including materials with independent magnetic and ferroelectric
subsystems, like some boracites, Aurivillius phases, hexagonal
manganites, and the lone-pair ferroelectrics with magnetic ions
\cite{khomskii}. Finally, in some perovskite manganites like
TbMnO$_3$ or GdMnO$_3$ it was proven experimentally
\cite{kimura,kimura05} that the onset of helical magnetic order
induces spontaneous ferroelectric (FE) polarization
\cite{kenzelmann,arima}. Dzyaloshinskii-Moriya type interactions
have been utilized to explain the ferroelectricity which is induced
by the helical spin structure \cite{katsura05,mostovoy, sergienko}.
A similar spin-driven ferroelectricity is believed to be operative
in Ni$_3$V$_2$O$_8$ \cite{lawes}.

The existence of an additional energy scale in ME compounds can lead
to the appearance of corresponding excitation of ME origin.
Recently, such excitations, called electromagnons, have been
observed experimentally \cite{nphys,sushkov} and it has been shown
that electromagnons are the relevant collective ME modes in these
materials. Electromagnons are strongly renormalized spin waves which
are coupled to optical phonons and can be excited by an ac electric
field. In TbMnO$_3$ and GdMnO$_3$ it has been documented that these
new excitations exist not only in the magnetic phase characterized
by the helical spin structure, but also in the longitudinally
modulated (sinusoidal) structure, provided that a "helical-type"
vector component of the spin-wave is dynamically induced via the ac
electric field \cite{nphys,gdfir}. The appearance of electromagnons
is supported by a theoretical modelling of elementary excitations in
helical magnets \cite{katsura} and by a polaron-like excitation
scheme of coupled phonons and magnons \cite{chupis}. In addition to
re-normalized phonons a new excitation has been predicted for the ME
state which originates from magnons and reveals a frequency
proportional to $\sqrt{SJD}$, where $S$ is the spin value, $J$ the
exchange coupling and $D$ the magnetic anisotropy. The coupling
between electromagnons and phonons has been verified experimentally
for GdMnO$_3$ \cite{gdfir} and for Eu$_{0.75}$Y$_{0.25}$MnO$_3$
\cite{aguilar}.

The analysis of the novel excitations in perovskite multiferroics
like TbMnO$_3$ still remains complicated due to the interplay
between the magnetic sublattices of the manganese and of the rare
earth ions. Further difficulties arise e.g. in GdMnO$_3$ as this
compound is close to a metastable ground state \cite{gdfir,jo1} and
can hardly be investigated using neutron scattering due to strong
absorption. Therefore, the details of the magnetic structure
especially in the ME-relevant incommensurate phases remains unknown.
To solve these problems, it is of interest to investigate materials
without the additional complexity of the rare earth magnetism. The
system Eu$_{1-x}$Y$_x$MnO$_3$ is a relevant magnetoelectric compound
without rare earth magnetism since Mn$^{3+}$ is the only magnetic
ion. In addition, yttrium doping allows a continuous tuning of the
magnetoelectric properties in this system and seems to increase the
strength of the ME coupling \cite{jo,jmmm}.

In this paper we present detailed investigations of the terahertz
spectra of yttrium-doped Eu$_{1-x}$Y$_x$MnO$_3$ in the concentration
range $0 \le x \le 0.5$. Because the characteristic energies of the
magnetoelectric contribution in perovskite manganites lie in the
terahertz frequency range, this region is especially important to
prove the existence of electromagnons and to study the spectral
changes of the ME contribution with doping.

\section{Experimental details}

Single crystals of Eu$_{1-x}$Y$_x$MnO$_3$ have been grown using the
floating-zone method with radiation heating. The samples were
characterized using X-ray, magnetic and dielectric measurements
\cite{jo}. The transmittance experiments at terahertz frequencies (3
cm$^{-1}$ $<\nu<$ 40  cm$^{-1}$) were carried out in a Mach-Zehnder
interferometer arrangement \cite{volkov} which allows measurements
of amplitude and phase shift in a geometry with controlled
polarization of the radiation. The absolute values of the complex
dielectric permittivity $\varepsilon^*=\varepsilon_1+i\varepsilon_2$
were determined directly from the measured spectra using the Fresnel
optical formulas for the complex transmission coefficient. The
experiments in external magnetic fields were performed in a
superconducting split-coil magnet with polypropylene windows
allowing to carry out transmittance and phase shift experiments in
magnetic fields up to 7\,T.

\begin{figure}[]
\includegraphics[width=6cm,clip]{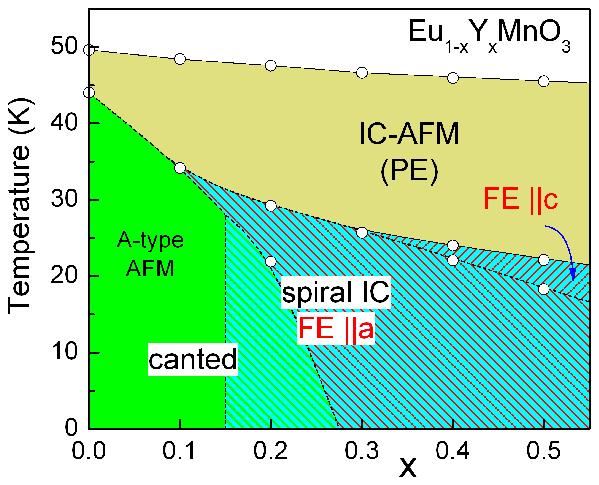}
\caption{(T-x) phase diagram of Eu$_{1-x}$Y$_x$MnO$_3$ reproduced
from Ref. \cite{jo}. The notation of different magnetic phases is
given on the basis of magnetization data. AFM - antiferromagnetic
phase, IC - incommensurate, PE - paraelectric, FE - ferroelectric.
Exact magnetic structures of different phases is still unknown and
is indicated in analogy to other perovskite multiferroics.}
\label{fphase}
\end{figure}

\section{Results}

Figure \ref{fphase} reproduces the phase diagram of Y-doped
EuMnO$_3$ from Ref. \cite{jo} which has been obtained using
structural, magnetic, dielectric and thermodynamic experiments. In
the doping range $0\le x \le 0.5$  Eu$_{1-x}$Y$_x$MnO$_3$
 orders antiferromagnetically between 45 and 50 K only slightly depending upon
the yttrium content. From the point of view of the magnetodielectric
effect (i.e. changes in dielectric permittivity by magnetic field)
and the observation of electromagnons, the phase diagram presented
can be divided into four regimes, which we will characterize
separately. i) In the low-doping range $0\le x \le 0.1$ in the
incommensurate (IC) antiferromagnetic (AFM) phase weak ME effects
are observed and the electromagnons are over-damped and not well
defined. The IC-AFM phase in this region is followed by the canted
(CA) antiferromagnetic phase which shows no magnetoelectric effect.
ii) For Y-doping around $x=0.2$ the electromagnons are clearly
observed in the spectra and can be suppressed by external magnetic
fields, which leads to strong magnetic field-dependence of the
dielectric permittivity. iii) For $x\approx 0.3$ the electromagnons
are strong in the ferroelectric (FE) phase at low temperatures but
they are not sensitive to external magnetic fields up to 7 T. iv)
The region $0.4\le x \le 0.5$ is closely similar to $x\approx 0.3$,
but here the dielectric permittivity is weakly dependent upon the
external magnetic field in the narrow temperature range of the
competition between ferroelectric phases with electric polarization
parallel to a- and c-axes, respectively.

Within the accuracy of the present experiments nonzero ME
contribution has been observed for $\tilde{e}||a$-axis only and no
effects could be detected for other crystallographic directions.
Therefore, only the a-axis results will be presented below.

\subsection{Weakly-magnetoelectric region ($x\le 0.1$)}

\begin{figure}[]
\includegraphics[width=6cm,clip]{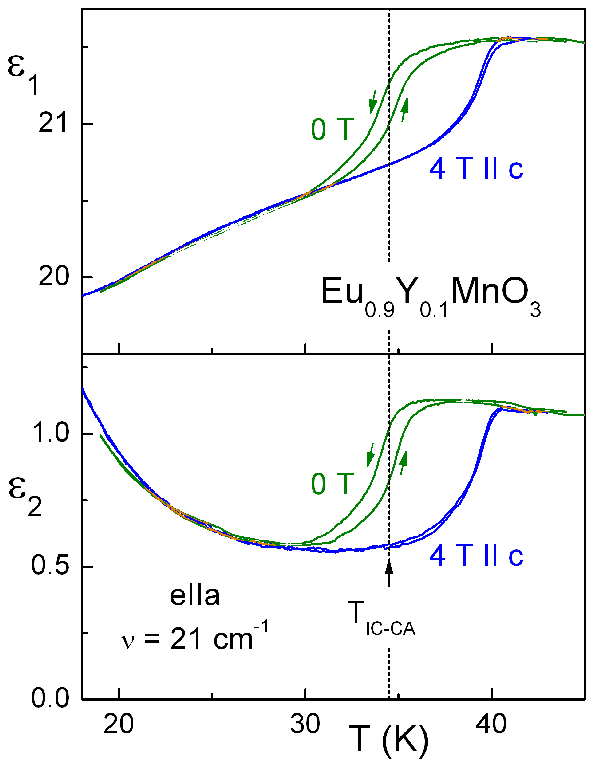}
\caption{Temperature dependence of the a-axis dielectric
permittivity of Eu$_{0.9}$Y$_{0.1}$MnO$_3$ in zero external field
(green) and at $\mu_0H=4$ T along the c-axis (blue). Upper panel -
real part, lower panel -imaginary part. $T_{IC-CA}$ indicates the
transition between the incommensurate and canted antiferromagnetic
states. This temperature has been obtained from the magnetization
experiments} \label{f10tem}
\end{figure}

The properties of Eu$_{0.9}$Y$_{0.1}$MnO$_3$ are representative for
this doping range and similar results have been obtained for pure
EuMnO$_3$ as well. Figure \ref{f10tem} shows the temperature
dependence of the dielectric permittivity of
Eu$_{0.9}$Y$_{0.1}$MnO$_3$ in zero magnetic field and in static
field of $\mu_0 H = 4$ T parallel to the c-axis. The steps in the
dielectric constant both in real and imaginary parts are observed
close to $T_{IC-CA}\simeq 34$ K. Higher dielectric constant and
stronger absorption in the high-temperature incommensurate magnetic
phase reflects the existence of additional contributions from the
magnetoelectric interactions. These additional contributions are
absent in the canted antiferromagnetic (CA-AFM) phase, which
explains lower dielectric constant below $T_{IC-CA} \simeq 34$ K.
The shift of the IC-CA phase transition by the static field is well
seen in this representation. Because the canted antiferromagnetic
phase contains a weak ferromagnetic component parallel to the
c-axis, the application of a strong external magnetic field in this
direction favors the CA-AFM phase and shifts the phase transition to
higher temperatures. No such effects have been observed for other
direction of the static magnetic field.

\begin{figure}[]
\includegraphics[width=6cm,clip]{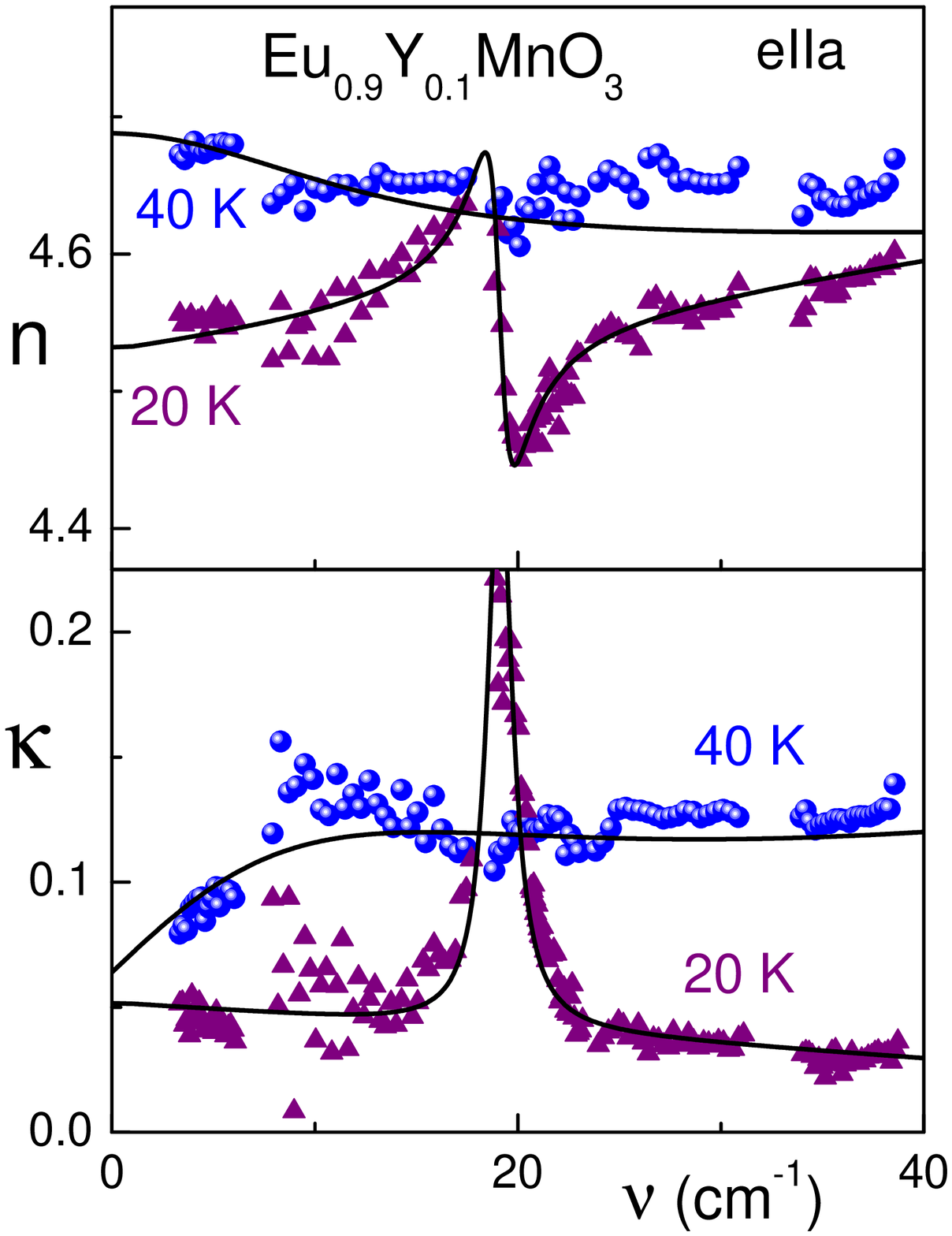}
\caption{Terahertz spectra of Eu$_{0.9}$Y$_{0.1}$MnO$_3$ in
incommensurate (40 K data) and canted (20 K data) antiferromagnetic
phases. Upper panel - refractive index, lower panel - absorption
coefficient. Narrow mode at $\nu \simeq 19$ cm$^{-1}$ represent the
antiferromagnetic resonance. Broad additional absorption for $T=40$
K is of magnetoelectric origin. Symbols represent experimental data,
lines show the fits using the sum of Lorentzians and a Debye
relaxator.} \label{f10frq}
\end{figure}

\begin{figure}[]
\includegraphics[width=6cm,clip]{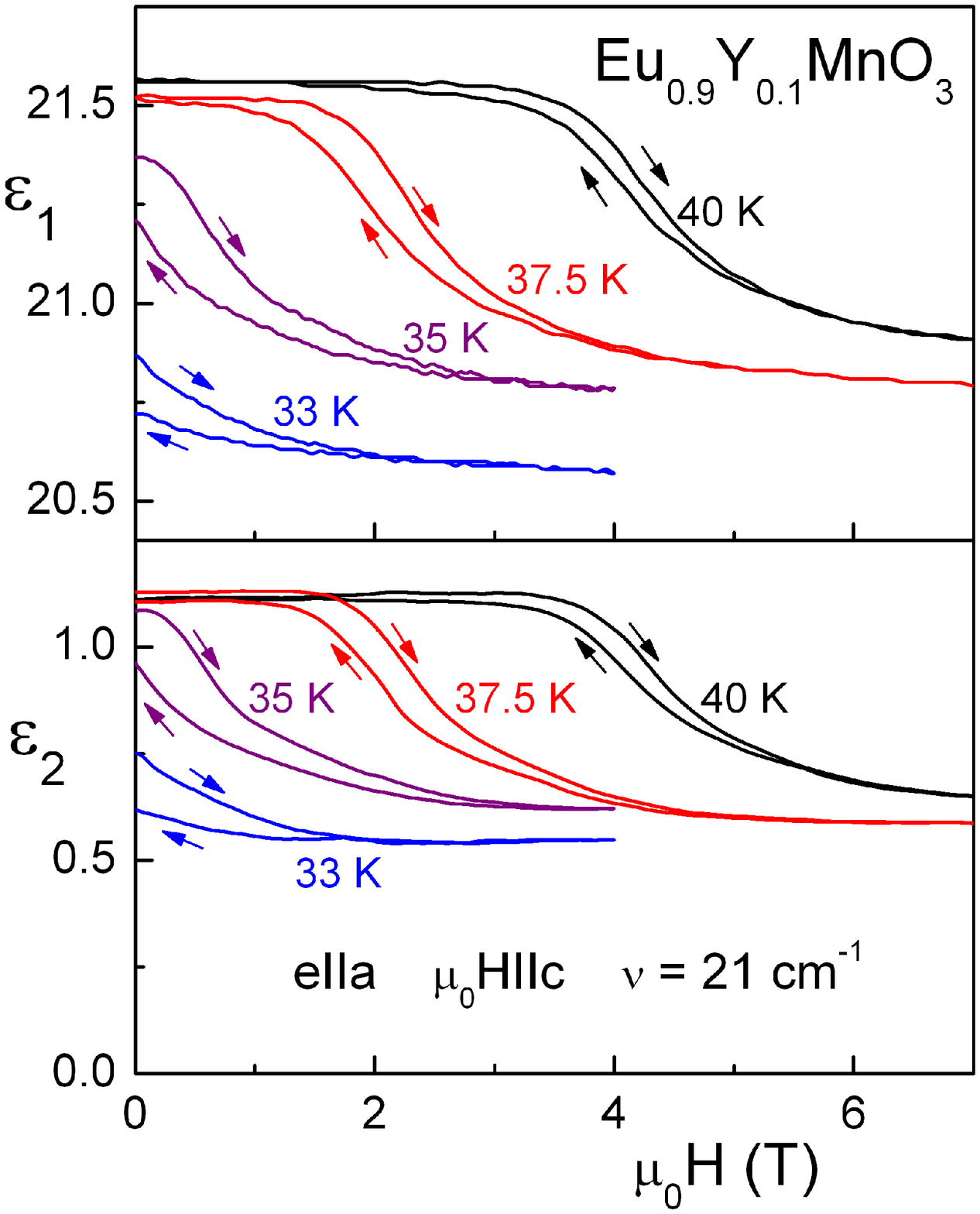}
\caption{Magnetic field dependence of dielectric permittivity of
Eu$_{0.9}$Y$_{0.1}$MnO$_3$ close to the IC-CA transition between two
antiferromagnetic phases. Upper panel shows the real part, lower
panel - imaginary part. } \label{f10fld}
\end{figure}

In analogy to the spectral properties of the magnetoelectric phases
in  GdMnO$_3$ and TbMnO$_3$ where weak but distinct electromagnons
exist already in the IC phase \cite{nphys}, we could expect the
existence of characteristic excitations (electromagnons), which
govern the magnetoelectric properties in ME materials. In order to
prove this similarity the terahertz spectra of
Eu$_{0.9}$Y$_{0.1}$MnO$_3$ have been measured in the IC and CA-AFM
phases. Typical results of these experiments are shown in Fig.
\ref{f10frq}. A strong and narrow absorption mode is observed in
these spectra close to 19 cm$^{-1}$ in the CA-AFM phase (20 K data).
This excitation represents the antiferromagnetic resonance (AFMR),
which for the studied polarization of ac magnetic field
($\tilde{h}||c$, b-cut sample) corresponds to the
quasiantiferromagnetic mode of the canted antiferromagnetic phases
\cite{afmr} and is well documented e.g. in La$_{1-x}$Sr$_x$MnO$_3$
\cite{lsmo}. The AFMR mode is of magnetic origin and therefore
cannot be plotted together with the spectra of dielectric
permittivity. Therefore, the $(n, \kappa)$ representation has been
chosen in this case, where $n+i\kappa = \sqrt{\varepsilon^* \mu^*}$.
However, far from the AFMR mode the response is purely dielectric
and the dielectric permittivity can be calculated as
$Re(\varepsilon)\simeq n^2$ and $Im(\varepsilon)\simeq 2n \kappa$.
In this approximation the upper panel of Fig. \ref{f10frq} reflects
qualitatively the behavior of $\varepsilon_1$ and the lower panel of
$\varepsilon_2$, respectively. Comparing the spectra at 20 K and at
40 K clear additional contribution can be seen in the
magnetoelectric IC-phase. This is seen both as a broad absorption in
the spectra of $\kappa(\nu)$, and as increased low-frequency values
of the refractive index in the IC-phase. Comparing this spectra to
that of GdMnO$_3$ and TbMnO$_3$ \cite{nphys}, this additional
contribution closely resembles electromagnon-like excitations.
However, in the case of Eu$_{0.9}$Y$_{0.1}$MnO$_3$ electromagnons
are not well-defined in energy and are seen as broad contribution
only. By fitting the spectra using a Debye-like relaxator (solid
lines in Fig. \ref{f10frq}, $T=40$ K), the characteristic frequency
of the ME-excitation in Eu$_{0.9}$Y$_{0.1}$MnO$_3$ can be estimated
as $\nu \sim 10$ cm$^{-1}$. Of course in this case this
characteristic frequency corresponds to an inverse mean lifetime of
the excitations.

In the vicinity of the IC-CA transition in
Eu$_{0.9}$Y$_{0.1}$MnO$_3$ the magnetoelectric contribution is
unstable against external magnetic fields along the c-axis, because
the application of the magnetic field in this direction favors the
CA-AFM phase. In this temperature range the ME effects can be
observed at terahertz frequencies. The results of ME experiments are
represented in Fig. \ref{f10fld} which shows real and imaginary
parts of the dielectric constant as function of external magnetic
field. In the temperature range 35 K $ \lesssim T \lesssim$ 40 K the
dielectric properties can be easily switched between the values
corresponding to IC and CA-phases. The observed changes in the
permittivity correspond well to the difference between the spectra
of both phases as shown in Fig. \ref{f10frq}.

\subsection{Intermediate doping range ($x \sim 0.2$) \label{sec20}}

\begin{figure}[]
\includegraphics[width=6cm,clip]{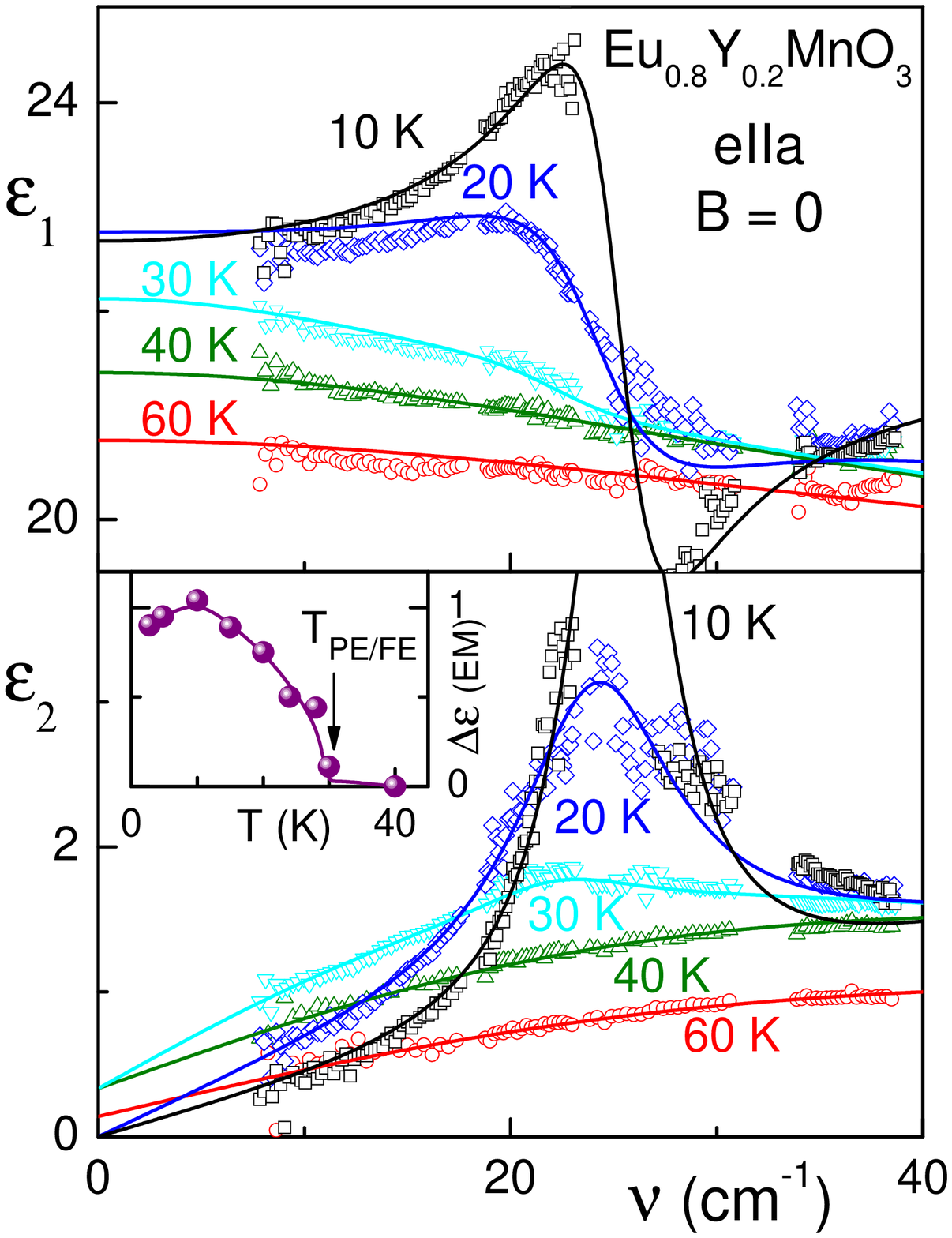}
\caption{Spectra of the dielectric permittivity along the a-axis of
Eu$_{0.8}$Y$_{0.2}$MnO$_3$ in the terahertz frequency range and
without external magnetic field. Upper panel - real part, lower
panel - imaginary part. Symbols - experiment, lines - Lorentzian
fit. Inset shows the dielectric contribution of electromagnon at
$\nu \simeq 24$ cm$^{-1}$.} \label{f20frq}
\end{figure}

For Eu$_{1-x}$Y$_{x}$MnO$_3$ being close to $x=0.2$ the collinear
spin order for $T < T_N$ is not ferroelectric, but for $T \le 29$ K
it is followed by a spiral spin structure which induces
ferroelectricity with the polarization $P||a$ \cite{jo}. It is
important to note that this spiral spin structure reveals a
ferromagnetic component, probably due to a conical-like distortion
\cite{jo}. Although the amplitude of the magnetoelectric effects in
this compound $\Delta \varepsilon_1 (H)$ is only slightly stronger
than in Eu$_{0.9}$Y$_{0.1}$MnO$_3$, electromagnons can be clearly
observed in the spectra and can be well fitted with a Lorentzian
with an eigenfrequency close to $\nu \simeq 24$ cm$^{-1}$ (Fig.
\ref{f20frq}). Figure \ref{f20frq} shows real and imaginary parts of
the dielectric spectra of Eu$_{0.8}$Y$_{0.2}$MnO$_3$ at different
temperatures and in zero external magnetic field. Already with the
onset of the IC collinear phase at $T \simeq 48$ K a broad terahertz
absorption starts to grow and marks an approaching of the FE state.
The dielectric spectra can be fitted using a Debye relaxator with
characteristic frequency of roughly 25 cm$^{-1}$. In addition to
this broad contribution, a well-defined electromagnon appears in the
FE state below $T=29$ K, which corresponds to the phase transition
from the paraelectric (PE) into the ferroelectric (FE) state. The
inset in the lower panel of Fig. \ref{f20frq} shows the dielectric
contribution of the electromagnon which exists in the ferroelectric
phase only. The growth of the spectral weight of the electromagnon
takes place on the costs of the relaxator that nevertheless survives
up to the lowest temperatures.

\begin{figure}[]
\includegraphics[width=6cm,clip]{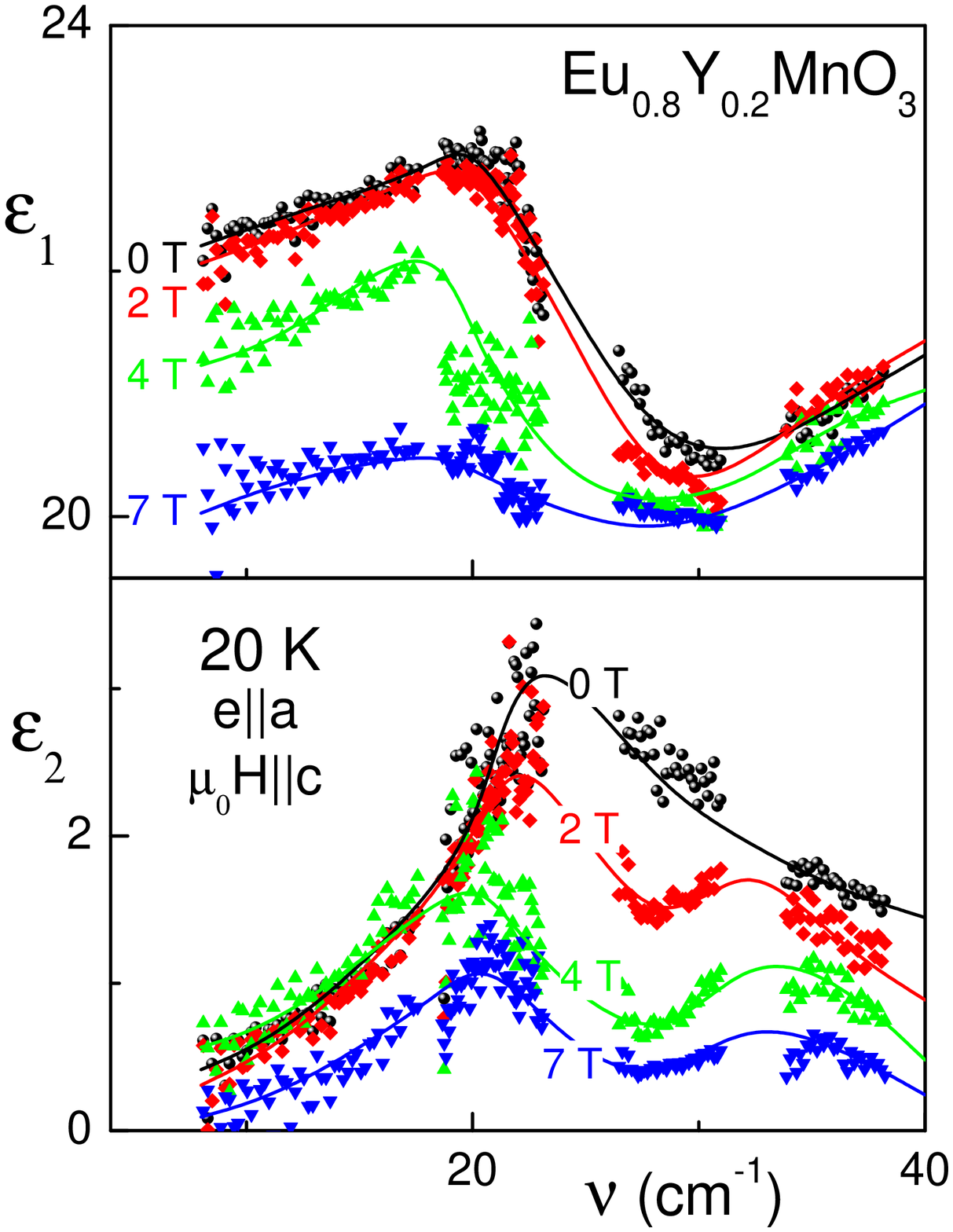}
\caption{Suppression and splitting of electromagnon in external
magnetic field in the a-axis dielectric spectra of
Eu$_{0.8}$Y$_{0.2}$MnO$_3$. Upper panel - real part, lower panel -
imaginary part. Symbols - experiment, lines - Lorentzian fit. }
\label{f20frfld}
\end{figure}

Eu$_{0.8}$Y$_{0.2}$MnO$_3$ reveals close similarities with the
magnetoelectric properties of GdMnO$_3$ and TbMnO$_3$. This
comprises characteristic frequency of electromagnon, typical values
of the dielectric permittivity, and includes the possibility to
suppress the electromagnon using external magnetic fields. Figure
\ref{f20frfld} shows the dielectric spectra of
Eu$_{0.8}$Y$_{0.2}$MnO$_3$ for different external magnetic fields
parallel to the c-axis. Compared to GdMnO$_3$ and TbMnO$_3$ the
suppression of the magnetoelectric contribution is more gradual and
is not fully finished even at $\mu_0H = 7$ T. This difference could
be due to the fact that the FE phase of Eu$_{0.8}$Y$_{0.2}$MnO$_3$
exhibits a ferromagnetic component and a different field dependence
compared to the pure antiferromagnetic spiral can be expected. In
addition, fine structure of the residual absorption is observed in
the spectra and can be approximated by two excitations at 21
cm$^{-1}$ and 34 cm$^{-1}$, respectively.

It seems that the spectra of the magnetoelectric perovskite
manganites cannot be described using just one single electromagnon
and furher components can be separated in the spectra. In addition
to the splitting of electromagnon observed in Fig.\ \ref{f20frfld},
the above-mentioned Debye-like contribution can be seen in  all
spectra of Eu$_{1-x}$Y$_{x}$MnO$_3$. This contribution was
dominating for Eu$_{0.9}$Y$_{0.1}$MnO$_3$ and in
Eu$_{0.8}$Y$_{0.2}$MnO$_3$ it gradually transfers its spectral
weight into electromagnons. Complicated spectra of electromagnons
have been observed in TbMnO$_3$ as well, both in dielectric
permittivity \cite{nphys} and in inelastic neutron scattering
experiments \cite{braden}. Finally, recent FIR experiments in
Eu$_{0.75}$Y$_{0.25}$MnO$_3$ \cite{aguilar} revealed the existence
of further electromagnon around $\nu_2=80$ cm$^{-1}$ in addition to
a low-frequency electromagnon at $\nu_1= 25$ cm$^{-1}$.

\subsection{Ferroelectric range ($x\geq 0.3$)}

\begin{figure}[]
\includegraphics[width=6cm,clip]{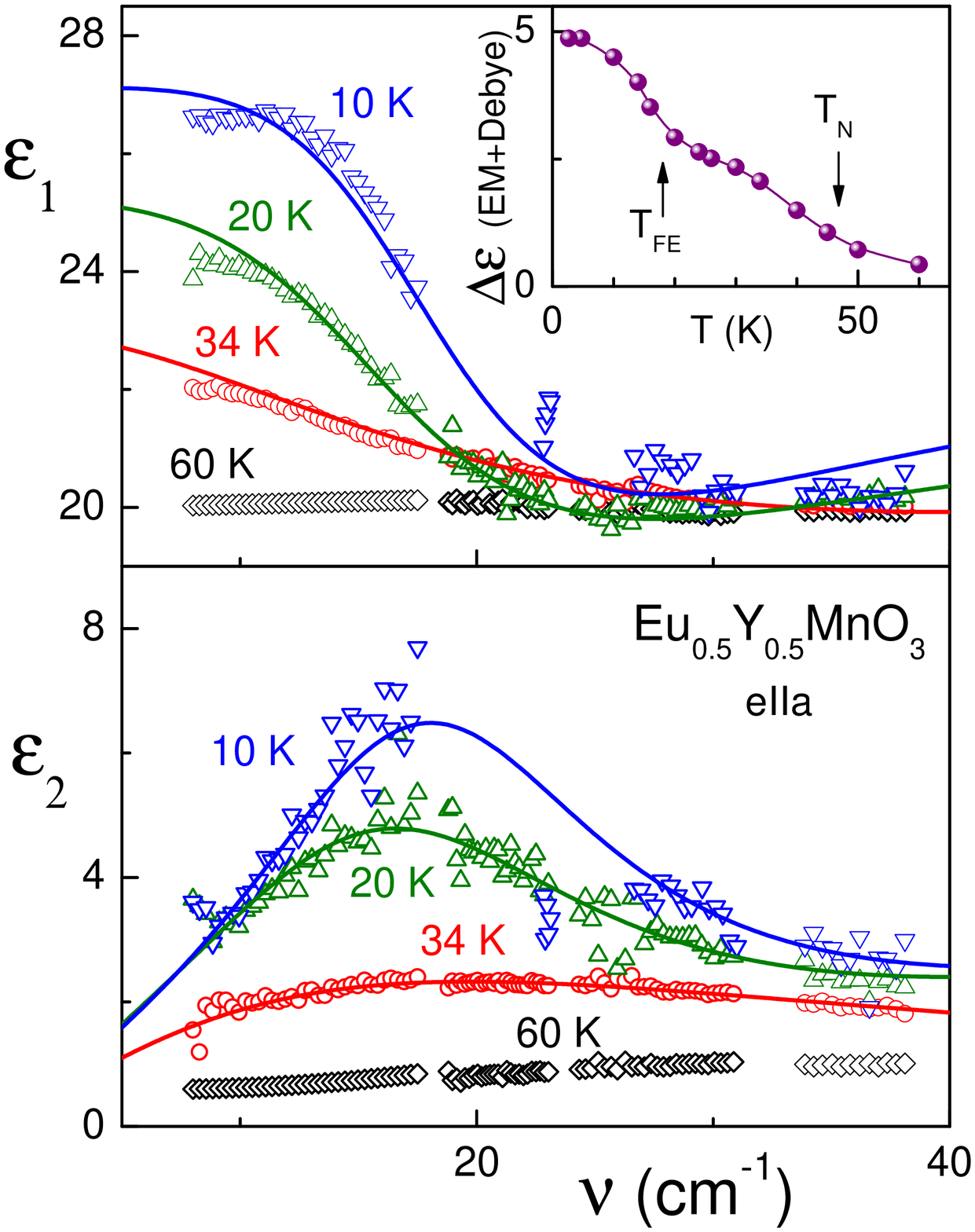}
\caption{Spectra of dielectric permittivity along the a-axis of
Eu$_{0.5}$Y$_{0.5}$MnO$_3$ in the terahertz frequency range. Upper
panel - real part, lower panel - imaginary part. Symbols -
experiment, lines - Lorentzian fit. Inset shows the dielectric
contribution of electromagnon and of Debye relaxator.}
\label{f50frq}
\end{figure}

\begin{figure}[]
\includegraphics[width=6cm,clip]{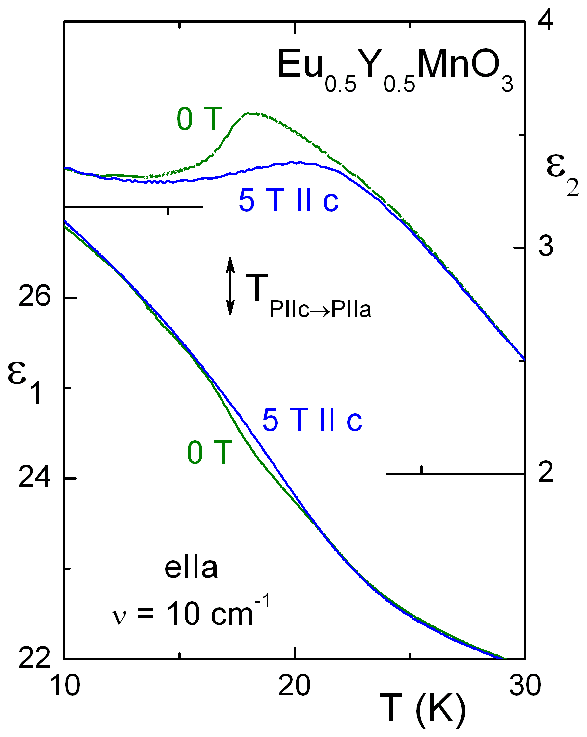}
\caption{Temperature dependence of the a-axis dielectric
permittivity of Eu$_{0.5}$Y$_{0.5}$MnO$_3$ in zero external field
and at $\mu_0H=5$ T along the c-axis. Upper panel - real part, lower
panel -imaginary part. Narrow region close to $T=18$ K, where the
permittivity is sensitive to magnetic field, corresponds to the
ferroelectric phase with $P||c$.} \label{f50tem}
\end{figure}

In this doping region and in zero external magnetic field the
permittivity spectra similar to Eu$_{0.8}$Y$_{0.2}$MnO$_3$ have been
observed. A typical example of these spectra is represented in Fig.
\ref{f50frq} which shows real and imaginary parts of the dielectric
permittivity of Eu$_{0.5}$Y$_{0.5}$MnO$_3$ for $\tilde{e} || a$.
Similar to other compositions, broad Debye-like contribution of ME
origin can be observed on entering the IC phase. This excitation is
getting more pronounced below $T_N \simeq 46$ K and its
characteristic damping frequency can be roughly identified as
$\Gamma \sim 20$ cm$^{-1}$. Temperature dependence of this
additional magnetoelectric contribution is shown in the inset to the
upper frame of Fig. \ref{f50frq}. Especially below the transition
from the paraelectric to the ferroelectric phase at $T_{PE/FE} \sim
20$ K this contribution transforms to a well-defined electromagnon
with a characteristic maximum in $\varepsilon_2$ and a step in
$\varepsilon_1$ close to 17 cm$^{-1}$ corresponding to its
eigenfrequency. The transition to the ferroelectric phase is also
seen as a change in slope in the temperature dependence of the
dielectric contribution of the electromagnon (inset to Fig.
\ref{f50frq}). We note that the dielectric contribution of the
electromagnons in this composition range is the strongest for the
Y-doping range investigated.

For the composition Eu$_{0.7}$Y$_{0.3}$MnO$_3$ no magnetic field
dependence of the dielectric properties could be observed for
external magnetic fields $\mu_0 H \le 7$ T within the experimental
accuracy. However, for $x \geq 0.4$ a narrow temperature region with
$P||c$ exists (Fig. \ref{fphase}) which can be influenced by
magnetic fields $\mu_0H||c$. Although this effect is extremely weak,
it can be observed in the temperature dependence of the dielectric
permittivity. Figure \ref{f50tem} shows the dielectric permittivity
of Eu$_{0.5}$Y$_{0.5}$MnO$_3$ at $\nu = 10$ cm$^{-1}$ as function of
temperature in zero external magnetic field and in a field of 5T
parallel to the c-axis. A clear hallmark of the phase transition to
the ferroelectric phase with $P||a$ is seen close to 17 K. This
transition is accompanied by a change of the rotation plane of the
spiral spin structure from the bc-plane (with $P||c$) to the
ab-plane (with $P||a$). In an external magnetic field with $\mu_0
H||c$ stabilizing the spiral structure with $P||a$, the onset of the
transition to this phase is shifted to higher temperatures up to
$T_{PE/FE}$. This results in a suppression the $P||c$ phase and in
weak magnetodielectric effects
($\Delta\varepsilon_1(H)/\varepsilon_1(0)\sim 0.8 \%$ in the real
part of the dielectric permittivity, Fig. \ref{f50tem}). We note
however that compared to the strong effects for low-doping
compositions the electromagnon contribution is rigid and remains
basically stable against the influence of external magnetic fields.

\section{discussion}

\begin{figure}[]
\includegraphics[width=5cm,clip]{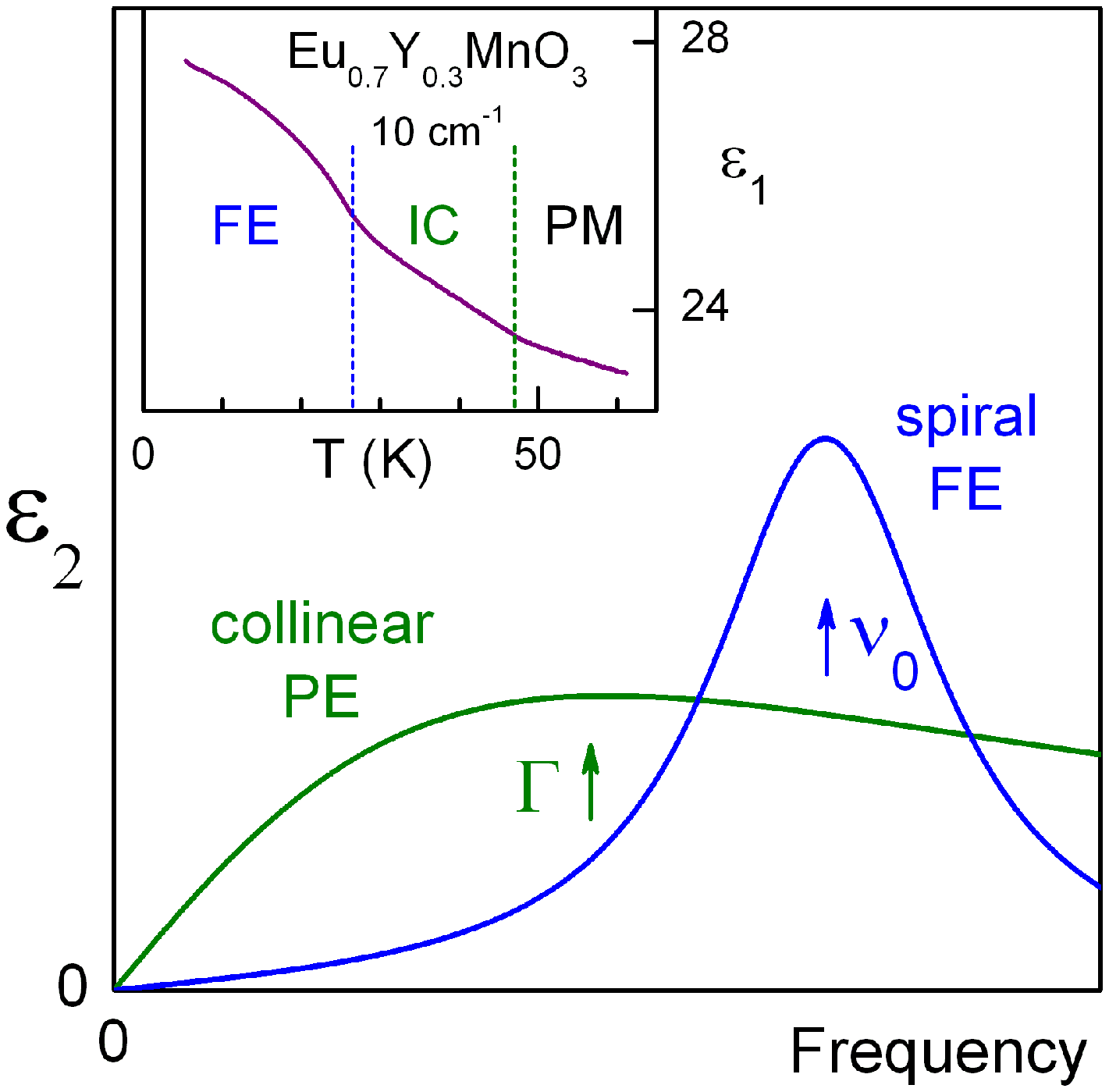}
\caption{Schematic representation of the terahertz spectra of
magnetoelectric manganites. Paramagnetic (PM) phase - no
magnetoelectric contribution, collinear incommensurate phase
(paraelectric (PE)) - broad Debye-like response, spiral
ferroelectric (FE) phase - well-defined electromagnons. $\Gamma$ is
the damping of the relaxator, $\nu_0$ is the eigenfrequency of the
electromagnon, and IC indicates the incommensurate antiferromagnetic
(collinear) phase. The inset shows the temperature dependence of the
low-frequency dielectric constant of Eu$_{0.7}$Y$_{0.3}$MnO$_3$ as
an illustration of the discussed behavior. Changes in slope in the
temperature dependence correspond well to the phase transitions to
the IC and FE states.} \label{fschema}
\end{figure}

Comparing the spectra of the dielectric permittivity of all
compositions investigated several similarities exist in the doping
series Eu$_{1-x}$Y$_{x}$MnO$_3$. General behavior of the terahertz
spectra is represented schematically in Fig. \ref{fschema}. Broad
magnetoelectric contribution exists in the terahertz spectra of the
dielectric permittivity in the collinear PE phase in all samples,
which seems to result from the same mechanisms as the electromagnon
responce. This contribution can be described using Debye-like
relaxation and is observed already in the paramagnetic phase
possibly due to magnon-like fluctuations.

In addition to the broad contribution a well defined electromagnon
starts to grow in the spiral FE state with a doping-dependent
resonant frequency which shifts from $\nu \simeq 25$ cm$^{-1}$ for
$x=0.2$ to $\nu \simeq 18$ cm$^{-1}$ for $x=0.5$. The dielectric
contribution of the electromagnon strongly depends upon
Y-concentration and grows from $\Delta\varepsilon \simeq 1$ for
$x=0.2$ to $\Delta\varepsilon \simeq 5$ for $x=0.5$. This reflects
the increase of the magnetoelectric coupling with the yttrium
doping. Indeed, for $x=0.2$ the stabilities of the ferroelectric and
the canted states are similar and both phases can be easily switched
by external magnetic fields. For $x \geq 0.3$ the FE state is stable
at low temperature and the electromagnons cannot be influenced by
external magnetic fields.

The interplay of different contribution to the dielectric
permittivity can be most clearly seen in the temperature dependence
of the low-frequency dielectric permittivity. According to the sum
rule \cite{dressel}
\begin{equation}
    \varepsilon_1(0)=1+ \frac{2}{\pi} \int_0^{\infty} \frac{\varepsilon_2(\omega)}{\omega} d \omega
\end{equation}
the low-frequency dielectric permittivity is a measure of all high
frequency contributions. The inset in Fig. \ref{fschema} shows the
temperature dependence of $\varepsilon_1$ in
Eu$_{0.7}$Y$_{0.3}$MnO$_3$ at $\nu=10$ cm$^{-1}$, i.e. below all
excitations observed in the terahertz spectra. On cooling the sample
through $T_N = 48$ K the increase in the slope of $\varepsilon_1(T)$
reflects the growth of the Debye-like excitation in the collinear PE
state. On further cooling, the spiral FE state is reached where
well-defined electromagnons are observed. The appearance of this
contribution is again observed as a change in slope of
$\varepsilon_1(T)$ at $T_{IC/FE}$.

Finally, the spectra of electromagnons reveal distinct fine
structure. This is most clearly documented in Fig. \ref{f20frfld}
and correlates well with other experimental observations both in
TbMnO$_3$ \cite{braden} and in Eu$_{0.75}$Y$_{0.25}$MnO$_3$
\cite{aguilar}.

\section{Conclusions}

Dielectric properties of yttrium-doped EuMnO$_3$ in the composition
range $0 \le x \le 0.5$ have been investigated in the terahertz-
frequency range. Nonzero magnetoelectric contribution to the
dielectric permittivity can be observed in all compositions for
$\tilde{e} ||a$ only. In the low doping range with coexisting
incommensurate and canted antiferromagnetic states ($x \le 0.2$) the
dielectric properties can be modified by external magnetic field
parallel to the c-axis especially on the border line between these
two phases. Well defined electromagnons are observed for $x \geq
0.2$ close to $\nu \sim 20$ cm$^{-1}$ and with strongly
doping-dependent dielectric strength. In addition to electromagnons,
a broad contribution of magnetoelectric origin is observed for all
compositions. Most naturally these contributions can be explained as
heavily overdamped electromagnons which already exist in the
collinear spin state.

This work has been partly supported by by DFG (SFB 484), and by RFBR
(04-02-16592, 06-02-17514).

\end{document}